\documentclass[prb,twocolumn,showpacs,superscriptaddress,amsmath,amssymb]{revtex4}

\usepackage{graphicx}%
\usepackage{dcolumn}%

\begin{document}

\title{Circular dichroism and bilayer splitting in the normal state of underdoped (Pb,Bi)$_2$Sr$_2$(Ca$_x$Y$_{1-x}$)Cu$_2$O$_{8+\delta}$ and overdoped (Pb,Bi)$_2$Sr$_2$CaCu$_2$O$_{8+\delta}$.}

\author{S. V. Borisenko}
\affiliation{Institute for Solid State Research, IFW-Dresden, P.O.Box 270116, D-01171 Dresden, Germany}

\author{A. A. Kordyuk}
\affiliation{Institute for Solid State Research, IFW-Dresden, P.O.Box 270116, D-01171 Dresden, Germany}
\affiliation{Institute of Metal Physics of National Academy of Sciences of Ukraine, 03142 Kyiv, Ukraine}

\author{S. Legner}
\author{T. K. Kim}
\author{M. Knupfer}
\author{C.~M. Schneider}
\author{J. Fink}
\affiliation{Institute for Solid State Research, IFW-Dresden, P.O.Box 270116, D-01171 Dresden, Germany}

\author{M. S. Golden}
\affiliation{Van der Waals-Zeeman Institute, University of Amsterdam, NL-1018 XE Amsterdam, The Netherlands}

\author{M. Sing}
\author{R. Claessen}
\affiliation{Experimentalphysik II, Universit\"at Augsburg, D-86135 Augsburg, Germany}

\author{A. Yaresko}
\affiliation{Max-Planck Institute for Physics of Complex Systems, D-01187 Dresden, Germany}

\author{H. Berger}
\affiliation{Institut de Physique de la Mat\'erie Complex, Ecole Politechnique F\'ederale de Lausanne, CH-1015 Lausanne, Switzerland}

\author{C. Grazioli}
\author{S. Turchini}
\affiliation{Istituto di Struttura della Materia, Consiglio Nazionale delle Ricerche, Area Science Park, I-34012 Trieste, Italy}

\date{\today}
\begin{abstract}

We report an ARPES investigation of the circular dichroism in the first Brillouin zone (BZ) of under- and overdoped Pb-Bi2212 samples. We show that the dichroism has opposite signs for bonding and antibonding components of the bilayer-split CuO-band and is antisymmetric with respect to reflections in both mirror planes parallel to the c-axis. Using this property of the energy and momentum intensity distributions we prove the existence of the bilayer splitting in the normal state of the underdoped compound and compare its value with the splitting in overdoped sample. In agreement with previous studies the magnitude of the interlayer coupling does not depend significantly on doping. We also discuss possible origins of the observed dichroism.

\end{abstract}

\maketitle

It is hard to overestimate the role of magnetic interactions in the high-temperature superconducting cuprates. While the spin fluctuations \cite{Chubukov} remain a reasonable candidate as a mediator of the pairing itself, the orbital currents form the basis of the approaches which seem to be promising in the explanation of the enigmatic pseudogap state.\cite{Chakravarty,Wen} Angle-resolved photoemission spectroscopy (ARPES) has proven to be a successful technique to study related feedback effects in the electronic spectrum.\cite{NormanPRL_97,Johnson,Gromko1,BorisPRL_03,KimPRL_03} To be directly sensitive to magnetism, ARPES experiment needs to resolve the spins of photoelectrons. Such an attractive opportunity to measure spin, momentum and energy simultaneously and with the same efficiency requires further improvement of the apparatus for "complete" photoemission - high spin resolution today still implies lower energy or/and momentum resolutions. To go around this problem one can suggest an alternative to a conventional explicit spin analysis. When using circularly polarized light as an excitation source, and in certain cases, the photoemission intensity itself may depend explicitly on the spin state of the excited electrons.\cite{OepenPRL,Garbe} Explained in terms of spin-dependent diffraction through the surface, this effect suggests a possibility to study spin-dependent processes with conventional spin integrating electron analyzers. On the other hand, circularly polarized excitation can also be used to test the time-reversal invariance of the electronic system which is broken in the presence of the orbital currents.\cite{Varma,Kaminski,Boris_condmat_03} Up to now, however, only a small part of the Brillouin zone of Bi$_2$Sr$_2$CaCu$_2$O$_{8+\delta}$ (Bi2212) in the close vicinity of the ($\pi$, 0)-point was the focus of attention in the studies using circularly polarized light.\cite{Kaminski,Boris_condmat_03} In addition, already the geometry of both experiments assumed a strong dichroic component, expected from a purely general consideration of the problem.\cite{Varma}

Here we extend our studies to the whole Brillouin zone. We reduce the "handedness" of our experiments, keeping the vectors of incidence, emission and the normal to the sample surface coplanar. In spite of this we clearly observe a strong (up to $\sim 18 \%$) dichroism which turns out to be of opposite sign for bonding and antibonding components and which is odd upon reflections in all mirror planes. We use this effect to prove the existence of the bilayer splitting in the normal state of an underdoped sample. This is in direct contradiction with recent claims that the bilayer splitting is absent already in the optimally doped Bi2212.\cite{Kaminski_noBS} Quantitative estimates give comparable results for the under- and overdoped samples implying the constancy of the bilayer splitting with doping in agreement with previous studies.\cite{Chuang,KordPRB_02}

The experiments were performed at the 4.2R beamline "Circular Polarization" at the ELETTRA storage ring using approximately 70$\%$ circularly polarized (CP) light from the elliptical wiggler-undulator. Spectra were collected in the angle-multiplexing mode of the SCIENTA SES-100 electron-energy analyzer. The overall average resolution in ({\bf k}, $\omega$)-space was set to 0.01 {\AA}$^{-1}$ x 0.02 {\AA}$^{-1}$ x 30 meV. An essential advantage of this experimental setup is that no mechanical movement is involved in the process of switching the helicity of the incoming radiation. Only the direction of the current in the coils of the wiggler-undulator needs to be reversed which takes approximately 30 seconds. This enables the successive recording of the spectra using the light of both polarizations with the other experimental parameters remaining unchanged. The excitation energy was $h\nu$=25 eV. High-quality single crystals of 5x1-superstructure-free, underdoped Y-Pb-Bi2212 ($T_{c}$=80 K) and overdoped ($T_{c}$=72 K) Pb-Bi2212 were mounted on the three-axis stepper motor driven cryo-manipulator allowing a precise (0.1$^{\circ}$) positioning of the sample with respect to the analyzer's entrance slit. Underdoping of the Pb-Bi2212 single crystals was done by partial substitution of calcium atoms by yttrium, which is known to be a good alternative to the de-oxygenation and is successfully investigated in different studies \cite{Y-doping}. Measurement temperatures of the overdoped and underdoped samples were 105 K and 120 K respectively. Fermi surface maps were recorded at room temperature from the {\it same} single crystals using mostly unpolarized light as reported earlier.\cite{KordPRB_02}

\begin{figure} [t!]
\includegraphics[width=8.47cm]{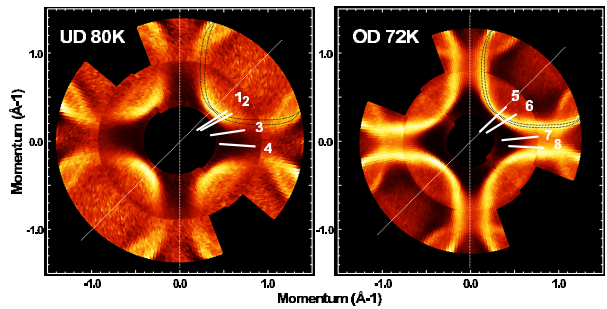} 
\caption{\label{Maps} Normal state Fermi surface maps of the UD80K and OD72K samples. White lines indicate the cuts in {\bf k}-space measured with circular light. Two mirror planes are shown as dashed white lines. Black dashed lines schematically show bilayer split components not clearly resolved at room temperature.}
\end{figure}

In Fig.~\ref{Maps} we present the results of the routinely performed measurements which are considered by us as an essential constituent of the sample characterization. Local maxima on the momentum distribution maps measured near the Fermi level correspond to the Fermi surface. The area of the Fermi surface seen in Fig.~\ref{Maps} as rounded squares directly corresponds to the number of charge carriers in the system, i.e. its doping level. One can easily notice the difference in sizes of the "barrels" even without any quantitative analysis. Moreover, the intensity near ($\pi$, 0)-point is much lower in case of the underdoped sample indicating the possible influence of the pseudogap. Comparison with the similar data for an optimally doped sample (not shown) together with these two observations clearly establishes the considerable difference in doping levels of the studied samples and their under- and overdoped characters respectively.\cite{KordPRB_02}

\begin{figure} [t!]
\includegraphics[width=8.6cm]{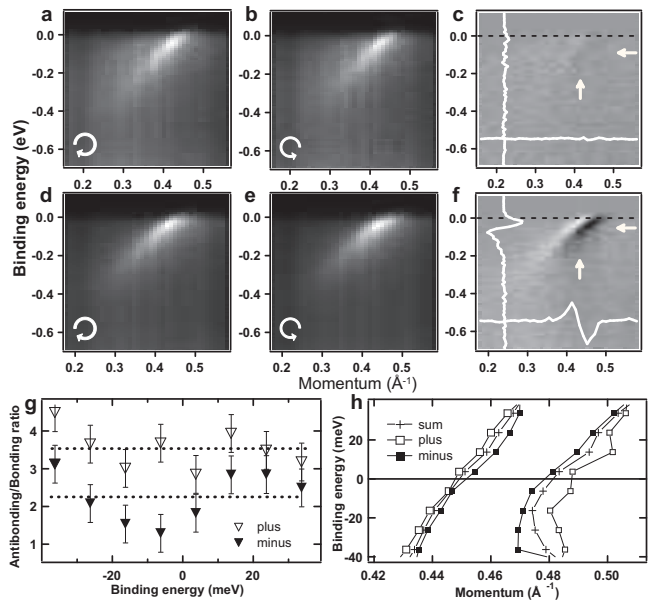}
\caption{\label{OD} Overdoped sample. EDMs taken with left- and right-hand circularly polarized light and their difference along the nodal (cut 5) direction (a-c) and along direction labelled 6 (d-f). White lines in panels (c) and (f) are the energy and momentum distributions of the dichroic signal obtained from the difference EDM by deciphering the grey scale along the cuts marked by white arrows. Results of the two-Lorentzian fitting procedure applied to the spectra shown in panels (d) and (e): ratio of the heights representing the relative bonding/antibonding contribution to the total spectral weight (g) and dispersion of the corresponding peak positions (h). Average values of the ratios are shown as black dashed lines (g).}
\end{figure}

Energy distribution maps (EDM) and their difference taken in the nodal direction (cut 5) of the overdoped sample using the CP light are shown in Fig.~\ref{OD}(a-c). A difference EDM is obtained by subtraction of the raw data shown in panel (b) from the data shown in panel (a) taking into account corresponding values of the storage ring current. The resultant grey scale image does not show any significant variations of the intensity upon the change of helicity. We note that the grey scale is the same for the panels (c) and (f) and ranges from -45000 counts (black) to +45000 counts (white) resulting in the grey background if the difference is negligible. Cut 6 crosses the Fermi surface along a direction which makes an angle of 11$^\circ$ with the $\Gamma$-($\pi$, $\pi$) line.
One expects here a crossing of two bands originating from the c-axis bilayer splitting, the latter having been observed now by a number of photoemission groups in overdoped samples. In this case the dichroic signal is remarkably strong ($\sim \pm8\%$) and has a well-defined structure (Fig.~\ref{OD} f). One would expect such a pattern if there are two closely separated dispersing features with the corresponding dichroism having the opposite sign for each of them. To check this we fitted the MDCs of panels (d), (e) and their sum with two Lorentzians. The fit appeared to be stable within the rather narrow energy interval ($\pm$ 40 meV) where an apparently asymmetric MDC's line shape was observed. The results are shown in Fig.~\ref{OD}(g, h). The extracted dispersions prove the presence of two bands split by 0.032(7) {\AA}$^{-1}$, though the position of the weaker (in terms of intensity) bonding component is defined with larger errors as seen from Fig.~\ref{OD}(h). More importantly, the ratio of the spectral weights corresponding to the bonding and antibondig bands is systematically higher for the spectra taken with right-hand circularly polarized light. The presented data thus confirm the selective character of the excitation from the bonding and antibonding bands when using light of different helicity. A qualitatively similar picture is observed for other cuts through the FS within the irreducible octant of the first BZ (not shown).

\begin{figure} 
\includegraphics[width=8.47cm]{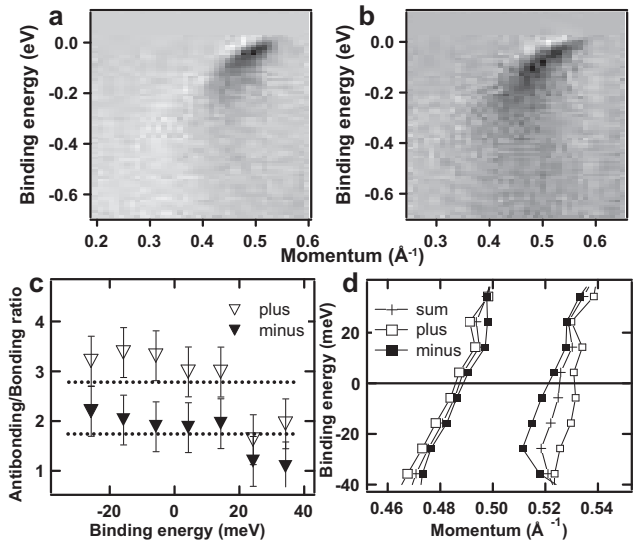}
\caption{\label{UD} Underdoped sample. (a, b) Difference EDM's along the directions 1 and 2. (c, d) Fitting results for the direction 1 equivalent to shown in Fig. 2 (g, h).}
\end{figure}

The absence of dramatic changes of this picture in the underdoped regime as compared to the overdoped case is evident from Fig.~\ref{UD}. There we show difference EDMs for the directions 1 and 2 (see Fig.~\ref{Maps}). The pattern is easily recognizable and qualitatively similar to the one shown in Fig.~\ref{OD}(f). Regardless of the origin of the dichroism, already on the basis of the visual comparison with the overdoped case, one can immediately conclude that the bilayer splitting is also present in the normal state of the underdoped sample, contrary to what is stated in Ref.\onlinecite{Kaminski_noBS}. Results of the fit for the direction 1 are nearly identical to those shown in Fig.~\ref{OD}: antibonding/bonding ratios are also of the order of 3 and 2 for the right- and left-hand CP light respectively and the size of the bilayer splitting is 0.037(5) {\AA}$^{-1}$ implying the constancy of this value with doping.\cite{Chuang} We note that the obtained values of the splitting in this direction of the BZ are larger than expected using the formula $t_{\perp} (\cos k_x a - \cos k_y a)^2 /2$ (Ref. \onlinecite{Andersen}) and estimated in Ref. \onlinecite{KordPRB_03} tight binding parameters . This is in agreement with our recent findings as for the finite nodal splitting.\cite{Kord_nodal}

\begin{figure} [t!]
\includegraphics[width=8.47cm]{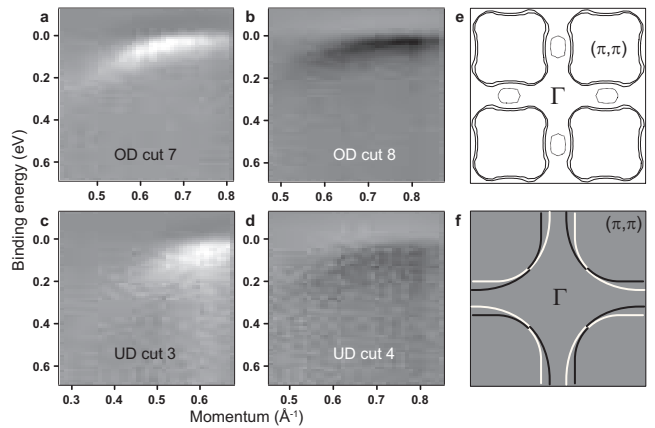}
\caption{\label{Dichro} Difference EDMs illustrating antisymmetric behavior of the dichroism with respect to reflection in the mirror plane for the overdoped (a, b) and underdoped (c, d) samples. Numbers of the cuts correspond to Fig.~\ref{Maps}. Grey scale ranges from -10$^5$ to 10$^5$ counts. (e) Fermi surface calculated within the LDA approximation including relativistic effects (spin-orbit interaction). (f) Schematic picture of the momentum distribution of the dichroism on the main Fermi surface in the first Brillouin zone.}
\end{figure}

Another characteristic feature of the momentum distribution of the dichroic signal is its antisymmetry with respect to the reflections in the crystal mirror planes. This is seen by comparing, for instance, the difference spectra presented in Fig.~\ref{Dichro} (a-d). The spectra taken with circularly polarized light of positive helicity on one side of the mirror plane (e.g. cut 7) essentially agree with the ones recorded using the light of negative helicity and measured on the other side (e.g. cut 8) and vice versa. As seen from the Fig.~\ref{Dichro} (a-d) this holds for both under- and overdoped samples. Qualitatively similar picture is observed for the $\Gamma$ - ($\pi$, $\pi$) mirror plane and at room temperature (not shown, see Ref.\onlinecite{Boris_condmat_03}).

Before going into the discussion of what could be the reason for such intensity variations we briefly summarize here three experimental results that follow from the presented and previously published data: (i) dichroism is observed everywhere in (\textbf{k}, $\omega$)-space except along the high symmetry lines, at different temperatures and in under- and overdoped samples of Pb-Bi2212; (ii) the sign of the dichroic signal changes upon reflection in all crystal mirror planes parallel to the c-axis; (iii) within the irreducible octant of the BZ the dichroism on the bonding and antibonding bands has the opposite sign. This is schematically illustrated in Fig.~\ref{Dichro} (f).

We consider three possible candidates which could be responsible for the dichroism in the studied samples. An interference effect, known in the literature as circular dichroism in the angular distribution (CDAD), can be observed in nonchiral systems only if the experimental setup has a definite chirality.\cite{Westphal} As mentioned above, our experimental arrangement is such that all three vectors (incidence, emission and normal to the surface) are always kept within the same plane. This rules out a purely geometric origin of the dichroism.\cite{Geometry}

The second possibility is the traditional magnetic circular dichroism in photoemission which requires the combination of spin-orbit coupling and exchange interaction in the initial or final state. It was demonstrated that the excitation with CP light under certain conditions results in {\it optical spin orientation} originating from the spin dependence of the dipole matrix elements.\cite{Woehlecke} The process becomes possible because the spin character of a given state is coupled to a wave function of certain spatial symmetry if one takes into account the spin-orbit interaction. In other words, the electrons excited from spin-orbit split bands could be spin-polarized in a final state due to relativistic symmetry selection rules when using circularly polarized light \cite{Orientation}. Further, to detect this spin-polarization using a conventional analyzer the diffraction through the surface should be spin-dependent. This seems to be the case in Bi2212 since exactly Bi-O layers are the topmost after cleavage and Bi atoms are heavy enough to act as a natural "spin-filter". Moreover, within this scenario also the observed mirror plane antisymmetry of the dichroism is robustly understood. Both the photon spin and the spin polarization are axial vectors, so that they behave like angular-momentum vectors under a mirror operation. Taking this into account it is easy to show \cite{Garbe} that such antisymmetry would mean the flipping of the vector of spin-polarization upon reversal of the helicity. This last condition, in turn, is fulfilled for {\it centrosymmetric} crystals assuming that the time reversal symmetry is not broken.\cite{Garbe} To clarify the interplay between the bilayer splitting and spin-orbit interaction we have carried out relativistic band structure calculations of Bi2212. The basal plane-projected Fermi surface is shown in Fig.~\ref{Dichro} (e). Inclusion of relativistic terms into the computational scheme results in the two split bands, but only near ($\pi$, 0) they are split by the spin-orbit interaction and only due to large admixture of the Bi-states which are thus actively involved in the formation of the Fermi surface. This, however, is in contradiction with the generally accepted views as for the non-metallicity of the Bi-O layers. The so called "Bi-O pockets" have never been experimentally observed. In addition, presented experimental results clearly suggest that the effect persists all along the Fermi surface.

The third option is that the observed dichroism could also be attributed to orbital currents flowing in real space and thus breaking time-reversal symmetry. Naturally, the symmetry of the dichroism in reciprocal space requires that the order imposed by the orbital currents has to adequately reflect this symmetry. This condition limits the number of possible patterns. For instance, the pattern of currents suggested in Ref.\onlinecite{Varma} implies the breaking of the reflection symmetry in the $\Gamma$ - ($\pi$, 0) mirror plane and requires a considerable value of the dichroism in this mirror plane, which is apparently not the case, also in the present study. On the other hand, the arrangement of elementary currents expected in the "d-density wave" approach \cite{Chakravarty} would be in agreement with our results. In order to account for the different sign of the dichroism for bonding and antibonding components one could order these currents in neighboring planes antiferromagnetically. In any case, our observations rule out a clear relation of the dichroism to the pseudogap regime as it was detected over significant portion of the phase diagram of Bi2212. This in turn reduces the chances of the orbital currents to be a solution of the given problem.

While at present it is not exactly clear what the origins of the reported dichroic effects in Bi2212 are, it definitely opens up a promising possibility to differentiate between the bonding and antibonding bands in future ARPES studies. Such a possibility seems to be urgently needed to help disentangle true many-body effects from mere features of the band structure.

We are grateful to H. Eschrig, A. Lichtenstein and N. Plakida for the stimulating discussions and to R. H\"ubel for technical support. We acknowledge the support of the European Community - Access to Research Infrastructure action of the Improving Human Potential Programme. HB is grateful to the Fonds National Suisse de la Recherche Scientifique, and MSG to FOM. MS and RC are supported by the DFG through SFB 484.


\begin{thebibliography}{}
\bibitem{Geometry} When this condition is not fulfilled it is easy to see a strong dichroic signal as in Refs.\onlinecite{Kaminski, Boris_condmat_03}.
\bibitem{Orientation} It is important to realize that the electrons are not spin-polarized in the initial state in the case of non-magnetic materials, and it is the excitation process goverened by relativistic symmetry selection rules that populates final states with a defined spin character. For instance, such selection rules for (100) surfaces of cubic crystals with normally incident circularly polarized light  predict 100 $\%$ spin-polarization in the final state and unambiguously connect the spin character with a particular excitation channel. See Ref.\protect\onlinecite{Woehlecke}.
\end{thebibliography}
\end{document}